\documentclass[aps,prl,preprint,showpacs,amsmath]{revtex4}

\usepackage{epsfig,bm}

\newcommand{\ba}{\begin{eqnarray}}
\newcommand{\ea}{\end{eqnarray}}
\newcommand{\bmath}{\begin{subequations}}
\newcommand{\emath}{\end{subequations}}
\newcommand{\ban}{\begin{eqnarray*}}
\newcommand{\ean}{\end{eqnarray*}}
\newcommand{\tl}{\tilde{\ell}}
\begin{document}

\title{Supersymmetric Patterns in the Pseudospin Spin and Coulomb Limits 
of the Dirac Equation with Scalar and Vector Potentials}
\author{A. Leviatan}
\affiliation{Racah Institute of Physics, The Hebrew University,
Jerusalem 91904, Israel}
\date{\today}
\begin{abstract}
We show that the Dirac equation in 3+1 dimensions 
gives rise to supersymmetric patterns when the scalar and 
vector potentials are (i) Coulombic with arbitrary strengths or (ii) when 
their sum or difference is a constant, leading to 
relativistic pseudospin and spin symmetries. 
The conserved quantities and 
the common intertwining relation responsible for such patterns are discussed.
\end {abstract}

\pacs{24.10.Jv, 11.30.Pb, 21.60.Cs, 24.80.+y }

\maketitle

The Dirac equation for spin $1/2$ particles plays a central role in the 
relativistic description of atoms, nuclei and hadrons. 
In atoms the relevant 
potentials felt by the electron (or muon in muonic atoms) are Coulombic
vector potentials. Relativistic mean fields in nuclei generated by 
meson exchanges \cite{wal86}, 
and quark confinement in hadrons \cite{gromes91} necessitate a mixture 
of Lorentz vector and scalar potentials. 
Recently symmetries of Dirac Hamiltonians with 
such Lorentz structure have been shown to be relevant for explaining 
the observed degeneracies of certain shell-model orbitals in nuclei 
(``pseudospin doublets'') \cite{gino97}, 
and the absence of quark spin-orbit splitting 
(``spin doublets'') \cite{page01}, 
as observed in heavy-light quark mesons. The goal of the 
current letter is to show that the degeneracy patterns and relations between 
wave functions implied by such relativistic symmetries resemble 
the patterns found in supersymmetric schemes. The underlying 
mechanism responsible for such properties will be examined. 
The feasibility of such a proposal gains support from the fact that 
Dirac Hamiltonians with selected external fields 
are known to be supersymmetric \cite{Thaller92}, 
{\it e.g.}, for a vector Coulomb potential \cite{suku85}.

Supersymmetric quantum mechanics (SUSYQM), initially proposed as a 
model for supersymmetry (SUSY) breaking in field theory \cite{witten81}, 
has by now developed into a field in its own right, with applications in 
diverse areas of physics \cite{junker96}. 
The essential ingredients of 
SUSYQM are the supersymmetric Hamiltonian  
${\cal H} = \left ({H_1\quad\atop 0}{0\atop H_2}\right ) = 
\left ({L^{\dagger}L\atop 0}{0\atop LL^{\dagger}}\right )$ 
and 
charges 
$Q_{-} = \left ({ 0\atop L}{ 0\atop 0}\right )$, 
$Q_{+} = \left ({ 0\atop 0}{ L^{\dagger}\atop 0}\right )$ 
which generate the supersymmetric algebra
$[{\cal H},Q_{\pm}] = \{Q_{\pm},Q_{\pm}\}=0$, 
$\{Q_{-},Q_{+}\}= {\cal H}$. 
The partner Hamiltonians $H_1$ and $H_2$ satisfy an 
intertwining relation,  
$LH_1 = H_2L$, 
where in one-dimension the transformation operator 
$L = \frac{d}{dx} + W(x)$ 
is a first-order Darboux transformation 
expressed in terms of a superpotential $W(x)$. The 
intertwining 
relation 
ensures that 
if $\Psi_1$ is an eigenstate of $H_1$, 
then also $\Psi_2=L\Psi_1$ is 
an eigenstate of $H_2$ with the same energy, 
unless $L\Psi_1$ vanishes or produces an unphysical state ({\it e.g.} 
non-normalizable). 
Consequently, as shown in Fig.~1(a), the SUSY partner Hamiltonians 
$H_1$ and $H_2$ are 
isospectral in the sense that their spectra 
consist of pair-wise degenerate levels 
with a possible non-degenerate single state in one sector (when 
the supersymmetry is exact). The wave functions of the degenerate levels 
are simply related in terms of $L$ and $L^{\dagger}$. 
Such characteristic features define 
a supersymmetric pattern. 
The 
intertwining relation 
ensures such properties 
for any pair of Hamiltonians not necessarily 
factorizable. We will continue to use the term ``supersymmetric pattern'' 
also in such circumstances. In what follows  
we focus the discussion on supersymmetric patterns 
obtained in selected Dirac Hamiltonians. 

The Dirac Hamiltonian, $H$, for a fermion of mass~$M$ 
moving in external scalar, $V_S$, and vector,
$V_V$, potentials is given by 
$H = \hat{\bm{\alpha}}\bm{\cdot p}
+ \hat{\beta} (M  + V_S) + V_V$, 
where~$\hat{\bm{\alpha}}$, 
$\hat{\beta}$ are the usual Dirac matrices \cite{mul85}
and we have set $\hbar = c =1$. 
When the potentials are spherically symmetric: $V_S=V_S(r)$, $V_V=V_V(r)$, 
the operator 
$\hat{K} = -\hat{\beta}\,(
\bm{\sigma\cdot\ell} + 1)$, 
(with $\bm{\sigma}$ the Pauli matrices and 
$\bm{\ell} = -i\bm{r}\times \bm{\nabla}$), 
commutes with $H$ and 
its non-zero integer eigenvalues $\kappa = \pm (j+1/2)$ are used to label 
the Dirac wave functions
$\Psi_{\kappa,\,m} = 
r^{-1}
( G_{\kappa} [\,Y_{\ell}\,\chi\,]^{(j)}_{m},
iF_{\kappa}[\,Y_{\ell^{\prime}}\,\chi\,]^{(j)}_{m})$. 
Here $G_{\kappa}(r)$ and $F_{\kappa}(r)$ 
are the radial wave functions of the upper and lower components 
respectively, $Y_{\ell}$ and $\chi$ are the spherical harmonic and 
spin function which are coupled to angular momentum $j$ with 
projection $m$.
The labels $\kappa = -(j+1/2)<0$ and $\ell^{\prime}= \ell +1$ hold 
for aligned spin 
$j=\ell+1/2$ ($s_{1/2}, p_{3/2}$, etc.), while $\kappa = (j+1/2)>0$ 
and $\ell^{\prime}= \ell -1$ hold 
for unaligned spin $j=\ell-1/2$ 
($p_{1/2},d_{3/2},$ etc.). 
Denoting the pair of radial wave functions by
\ba
\Phi_{\kappa} = \left (
{G_{\kappa}\atop F_{\kappa}}
\right ) ~,
\label{radialwf}
\ea
the radial Dirac equations can be cast in Hamiltonian form, 
$H_{\kappa}\,\Phi_{\kappa} = E\,\Phi_{\kappa}$, 
with
\bmath
\ba
H_{\kappa} &=& 
\left (
\begin{array}{ll}
M + \Delta   & \quad -\frac{d}{dr} + \frac{\kappa}{r} \\
\frac{d}{dr} + \frac{\kappa}{r} & \quad -(M + \Sigma) 
\end{array}
\right )
\label{hk}\\
\Delta(r) &=& V_S + V_V\;\ , \;\; \Sigma(r) = V_S - V_V.
\label{delsig}
\ea
\label{hkgen}
\emath
We now look for 
Dirac Hamiltonians  
$H_{\kappa_1}$ and $H_{\kappa_2}$ 
which satisfy an intertwining relation of the form 
\ba
LH_{\kappa_1} = H_{\kappa_2}L ~.
\label{lh1h2}
\ea
Following \cite{deber02} we consider a matricial Darboux transformation 
operator
\ba
L = A(r)\frac{d}{dr} + B(r) 
\label{darboux}
\ea
where $A$ and $B$ are $2\times 2$ matrices with $r$-dependent entries 
$A_{ij}(r)$, $B_{ij}(r)$.
Relations~(\ref{lh1h2}) and~(\ref{darboux}) should be regarded as 
a system of equations for the unknown operator $L$ and the 
so-far unspecified potentials in 
$H_{\kappa}$ (\ref{hkgen}). 
The matrices $A(r)$ and $B(r)$ are found to be
\bmath
\ba
A_{11} &=& A_{22} =  a \;\; , \;\; A_{12} = -A_{21} = b \\
B_{11} &=& -b(M+\Delta) 
-\frac{1}{2r}a\,\omega_{-}(\omega_{+}+1)
+\frac{1}{2}c_2 \\
B_{22} &=& b(M+\Sigma) 
-\frac{1}{2r}a\, \omega_{-}\, (\omega_{+} - 1) +\frac{1}{2}c_2 \\
B_{12} &=& 
a V_V 
-\frac{1}{2r}b\, \omega_{+}(\omega_{-} + 1)
+\frac{1}{2}c_1 \\
B_{21} &=& 
-a V_V 
+\frac{1}{2r}b\, \omega_{+}\, (\omega_{-}-1) - \frac{1}{2}c_1
\ea
\label{ABmat}
\emath
where $\omega_{+}=(\kappa_1+\kappa_2)$, 
$\omega_{-}=(\kappa_2 - \kappa_1)$ and $a,b,c_1,c_2$ are constants. 
In addition, the following relations have to be obeyed
\bmath
\ba
&&
2a \left [ \, V_{S}^{\prime} 
+ V_V\frac{\omega_{+}}{r}\, \right ] 
-b\,\frac{\omega_{-}(\omega_{+}+1)(\omega_{+}-1)}{r^2} 
+ c_{1}\,\frac{\omega_{+}}{r} = 0 ~,\\
&&
a \left [\, -4V_{V}(M + V_{S}) + 
\frac{\omega_{+}(\omega_{-}+1)(\omega_{-}-1)}{r^2} \, \right ] 
+\, 2b\, \frac{\omega_{-}}{r}\, 
\left [\, \omega_{+}\,(M+V_{S}) +V_{V}\, \right ]
\nonumber\\
&& \qquad\quad 
- c_{2}\,\frac{\omega_{-}}{r} - 2c_{1}\,(M + V_{S}) = 0 ~, 
\qquad
\ea
\label{eqvsvs}
\emath
with $V_{S}^{\prime}$ denotes differentiation with 
respect to $r$. In the usual application of SUSYQM, one starts from 
a solvable Hamiltonian $H_1$ and uses 
the intertwining relation to obtain 
a new solvable Hamiltonian $H_2$. In the present case we employ a different 
strategy, namely, insist that both partner Hamiltonians $H_{\kappa_1}$ and 
$H_{\kappa_2}$ be of the 
form prescribed in Eq.~(\ref{hkgen}) with the {\it same} potentials, 
and look for solutions of Eq.~(\ref{eqvsvs}) such that the potentials 
are independent of $\kappa$. 
We find that there are six such solutions characterized by 
$\omega_{+},\omega_{-}=0,\pm 1$. 
The solution with $\omega_{-}=0$ is trivial ($\kappa_1=\kappa_2$), 
$L=-b\,H_{\kappa} + \frac{1}{2}c_2I$ where $I$ is the $2\times 2$ 
unit matrix.
The solutions with $\omega_{-}=\pm 1$ lead to constant potentials 
$V_S = S_0$ and $V_V = V_0$. The physically interesting 
solutions require $\omega_{+}=0,1,-1$ and lead to the Coulomb, 
pseudospin and spin limits respectively.

{\it The Coulomb limit} $(\kappa_1+\kappa_2=0)$\\ 
The solutions of Eq.~(\ref{eqvsvs}) with $\omega_{+}=\kappa_1+\kappa_2=0$ 
fix the potentials to be of Coulomb type 
\ba
V_S &=& \frac{\alpha_{S}}{r} + S_0 \;\; , \;\;
V_V = \frac{\alpha_{V}}{r} + V_0 ~, 
\label{coulmbpot}
\ea
with arbitrary strengths, $\alpha_S$, $\alpha_V$. 
The constants $S_0$ and $V_0$ amount to constant shifts in the mass 
and Hamiltonian respectively, and
henceforth will be omitted. 
In terms of the quantities
$\eta_1 = (\alpha_S M + \alpha_V E)/\lambda \,$, 
$\eta_2 = (\alpha_S E + \alpha_V M)/\lambda\,$, 
$\lambda = \sqrt{M^2 - E^2}\,$, 
$\gamma = \sqrt{\kappa^2 + \alpha_{S}^2 - \alpha_{V}^2}\,$, 
the quantization condition reads: 
$\gamma + \eta_1 = -n_r$ 
$(n_r=0,1,2,\ldots\,)$, 
and leads to the 
eigenvalues~\cite{mul85} 
$E^{(\pm)}_{n_r,\kappa}/M =
[\, -\alpha_{S}\alpha_{V} 
\pm \xi\sqrt{\,\xi^2 -\alpha_{S}^2 + \alpha_{V}^2}\,]/
(\alpha_{V}^2 + \xi^2)$, 
where $\xi= (n_r+\gamma)$, 
and the $\kappa$-dependence enters through the factor $\gamma$. 
The spectrum consists of two branches denoted by superscripts 
$(+)$ and $(-)$. 
The corresponding eigenfunctions are
\ba
\Phi_{n_r,\kappa} 
\propto 
\left ({ -\sqrt{M+E}
[(\kappa+\eta_2)F_1 + n_r F_2]\atop 
\sqrt{M-E}\,
[(\kappa+\eta_2)F_1 - n_r F_2\,]}
\right )
\rho^{\gamma}e^{-\rho/2} \;
\label{coulombwf}
\ea
where $E = E^{(\pm)}_{n_r,\kappa}$ 
and $F_1=F(-n_r,2\gamma+1,\rho)$, $F_2 = F(-n_r+1,2\gamma+1,\rho)$ 
are confluent hypergeometric functions in the variable $\rho = 2\lambda r$. 
The states and energies in each branch are labeled by $(n_r,\kappa)$. 
It is also possible to express the results in terms of the principal quantum 
number $N$ defined as $N= n_r + \vert \kappa\vert$, $(N=1,2,\ldots\,)$. 
For $n_r\geq 1$ 
the eigenvalues in each 
branch are two-fold degenerate with respect to the sign of 
$\kappa$, {\it i.e.} 
$E^{(+)}_{n_r,\kappa}=E^{(+)}_{n_r,-\kappa}$ and 
$E^{(-)}_{n_r,\kappa}=E^{(-)}_{n_r,-\kappa}$. 
For $n_r=0$ there is only {\it one} acceptable state for each $\kappa$, 
which has $\kappa<0$ for the $(+)$ branch and 
$\kappa >0$ for the $(-)$ branch. Equivalently, 
for a fixed principal quantum number $N$, 
the allowed values of $\kappa$ 
are $\kappa = \pm 1, \pm 2,\ldots, \pm (N-1),-N$ for the 
$(+)$ branch and $\kappa = \pm 1, \pm 2,\ldots, \pm (N-1),+N$ for the 
$(-)$ branch of the spectrum.

Focusing on the set of states with $\kappa_1=-\kappa_2\equiv \kappa$, 
the levels are separated according to the value of 
$\vert\kappa\vert = j+1/2$. 
For fixed $\kappa$, $E^{(+)}_{n_r,\kappa}$ is an increasing function 
of $n_r$ and, as shown in Fig.~1(b), 
for each value of $j$ we have a characteristic supersymmetric pattern.
There are two 
towers of energy levels, 
one for $-\vert\kappa\vert$ (with $n_r=0,1,2,\dots$) and one for 
$+\vert\kappa\vert$ (with $n_r=1,2,\ldots$). 
The two towers are identical, except that the 
$E^{(+)}_{n_r=0,-\vert\kappa\vert}$ 
level at the bottom of the $-\vert \kappa\vert$ 
tower is non-degenerate. 
Similar patterns of pair-wise degenerate levels with $\pm\kappa$ 
appear also in the $(-)$ branch of the spectrum. 
However, since for fixed $\kappa$, 
$E^{(-)}_{n_r,\kappa}$ is a decreasing function of $n_r$, 
the non-degenerate $E^{(-)}_{n_r=0,\vert\kappa\vert}$ level is now 
at the top of the $+\vert \kappa\vert$ tower, 
resulting in an inverted supersymmetric pattern. 
From Eqs.~(\ref{ABmat})-(\ref{eqvsvs}) 
we find the transformation operator to be
\ba
L = a \left (
\begin{array}{cc}
\frac{d}{dr} +\frac{\epsilon_{+}}{r} + \frac{M\alpha_{+}}{\kappa_1}  &  
-\frac{\alpha_S}{\kappa_1}\frac{d}{dr} + \frac{\alpha_V}{r} \\
\frac{\alpha_S}{\kappa_1}\frac{d}{dr} - \frac{\alpha_V}{r} &
\frac{d}{dr} -\frac{\epsilon_{-}}{r} - \frac{M\alpha_{-}}{\kappa_1}
\end{array}
\right ) ~, 
\label{Lcoulomb}
\ea 
where $\epsilon_{\pm} = \kappa_1 
+ \alpha_S\alpha_{\pm}/\kappa_1$ 
and $\alpha_{\pm}= (\alpha_S \pm \alpha_V)$. 
The operator $L$ connects degenerate 
states with 
$(n_r\geq 1,\pm\kappa)$, 
and annihilates the non-degenerate states with $n_r=0$
\ba
L\,\Phi_{n_r,\kappa_1} = C\,\Phi_{n_r,\kappa_2} 
\quad (\kappa_1=-\kappa_2) ~.
\label{LGF}
\ea
Here $C =\frac{a\lambda}{\kappa_1}\sqrt{n_r(\gamma -\eta_1)}$ 
and $\Phi_{n_r,\kappa}$ are given in Eq.~(\ref{coulombwf}). 
Constructing supersymmetric charges $Q_{\pm}$ and Hamiltonian 
${\cal H}$ from $L$ and $H_{\kappa_1}$, $H_{\kappa_2}$ in the manner 
described at the beginning of the letter, 
ensures that 
$[{\cal H},Q_{\pm}]=\{Q_{\pm},Q_{\pm}\}=0$, but now 
$\{Q_{-},Q_{+}\} \propto 
({\cal H} - E^{(+)}_{n_r=0,\kappa})
({\cal H} - E^{(-)}_{n_r=0,\kappa})$, 
with $\kappa=\kappa_1=-\kappa_2$. 
These relations represent a quadratic deformation 
of the conventional supersymmetric algebra \cite{deber02}, 
which arises because 
both the Dirac Hamiltonian and the transformation operator $L$ 
are of first order. 

The explicit solvability and observed degeneracies of the relativistic 
Coulomb problem are related to the existence of an additional conserved 
Hermitian operator 
\ba
B= -i\hat{K}\gamma_5\,\left (H - \hat{\beta}M\right ) + 
\frac{\bm{\sigma \cdot r}}{r} 
\left 
(\,\alpha_{V} M + \alpha_{S} H\,\right ) ~, \;\;\; 
\label{JL}
\ea
which commutes with the full Dirac scalar and vector Coulomb Hamiltonian, 
$H$, but anticommutes with $\hat{K}$.
This operator is a generalization of the 
Johnson-Lippmann operator   
applicable for $\alpha_S=0$ \cite{john50}. 

{\it The pseudospin limit} $(\kappa_1+\kappa_2 =1)$\\
The solutions of Eq.~(\ref{eqvsvs}) with $\omega_{+}=\kappa_1+\kappa_2=1$ 
require that the sum of scalar and vector potentials is a constant 
\ba
\Delta(r) = V_{S}(r) + V_{V}(r) = \Delta_0 ~.
\label{pspot}
\ea 
Under such condition the Dirac Hamiltonian is invariant under an SU(2) 
algebra, whose generators are \cite{bell75,ginolev98} 
\ba
{\hat{\tilde {S}}}_{\mu} =
\left (
\begin{array}{cc}
\hat {\tilde s}_{\mu} &  0 \\
0 & {\hat s}_{\mu}
\end{array}
\right ) ~.
\label{Sgen}
\ea
Here 
${\hat s}_{\mu} = \sigma_{\mu}/2$ are the usual spin 
operators, $\hat {\tilde s}_{\mu}= U_p {\hat s}_{\mu} U_p$ 
and $U_p = \frac{\bm{\sigma\cdot p}}{p}$. 
This relativistic symmetry has been used \cite{gino97} 
to explain the occurrence 
of pseudospin doublets in nuclei \cite{arima69}. 
The latter refer to the empirical observation 
of quasi-degenerate pairs of certain normal-parity
shell-model orbitals with non-relativistic 
single-nucleon radial, orbital, and total angular 
momentum quantum numbers: 
$(n,\ell,j = \ell + 1/2)$ 
and 
$(n-1,\ell + 2,j = \ell + 3/2)$. 
The doublet structure 
is expressed in terms of a ``pseudo'' orbital angular momentum, 
$\tilde{\ell}$ = $\ell+1$, and ``pseudo'' spin, $\tilde{s} = 1/2$, which 
are coupled to 
$j = \tilde{\ell}\pm \tilde s$. 
For example, $(n s_{1/2},(n-1)\,d_{3/2})$ will have
$\tilde{\ell}= 1$, etc. 
Such doublets 
play a central role in explaining features of nuclei \cite{bohr82}, 
including superdeformation 
and identical bands. 
In a relativistic description of nuclei \cite{wal86}, 
these non-relativistic wave functions 
are identified with the upper components of 
Dirac wave functions, $\Psi_{\kappa_1<0,m}$ and $\Psi_{\kappa_2>0,m}$ 
with $\kappa_1+\kappa_2=1$, 
which are eigenstates of a Dirac Hamiltonian with scalar and 
vector mean field potentials, approximately satisfying 
condition (\ref{pspot}). The corresponding lower components 
have $n$ nodes \cite{levgino01} and orbital angular momentum equal 
to $\tl$ \cite{gino97}. 
In the pseudospin limit these two Dirac states 
form a degenerate doublet, 
and their radial components satisfy 
$F_{\kappa_1} = F_{\kappa_2}$, and 
$\frac{dG_{\kappa_1}}{dr} + \frac{\kappa_1}{r}G_{\kappa_1} 
= \frac{dG_{\kappa_2}}{dr} + \frac{\kappa_2}{r}G_{\kappa_2}$.
These 
relations have been tested in numerous 
realistic mean field calculations in a variety of nuclei, 
and were found to be obeyed 
to a good approximation, especially for doublets near the Fermi 
surface \cite{ginomad98,ginolev01}. 
For potentials with asymptotic behavior as encountered in nuclei, 
the Dirac eigenstates for which both the 
upper ($G_{\kappa}$) and lower ($F_{\kappa}$) components have no nodes, 
can occur only for $\kappa <0$, and hence do not have a degenerate 
partner eigenstate (with $\kappa >0$) \cite{levgino01}. 
These nodeless Dirac states correspond to the shell-model states 
with $(n=0,\ell,j=\ell+1/2)$. 
For heavy nuclei such states with 
large $j$, {\it i.e.}, $0g_{9/2},\;0h_{11/2},\;0i_{13/2}$, 
are the ``intruder'' abnormal-parity states which, indeed, empirically 
are found not to be part of a doublet \cite{bohr82}. 
Altogether, as shown in Fig.~1(c), the ensemble of Dirac 
states with $\kappa_1+\kappa_2=1$ 
exhibits a supersymmetric pattern of twin towers with pair-wise degenerate 
pseudospin doublets sharing a common $\tl$, and an additional 
non-degenerate nodeless state at the bottom of the $\kappa_1<0$ tower. 
An exception to this rule 
is the tower with $\kappa_2=1$ ($p_{1/2}$ states with $\tl=0$), 
which is on its own, because states with $\kappa_1=0$ do not exist. 
From Eq.~(\ref{ABmat})-(\ref{eqvsvs}) 
we find the transformation operator to be
\ba
L = b \left (
\begin{array}{cc}
0 \quad & \frac{d}{dr} - \frac{\kappa_2}{r} \\
-\frac{d}{dr} - \frac{\kappa_1}{r} \quad & 
(2M + \Sigma + \Delta_0)
\end{array}
\right ) ~.
\label{Ldel0}
\ea
$L$ connects the two doublet states 
as in Eq.~(\ref{LGF}) but with $\kappa_1+\kappa_2=1$ 
and $C = b(M + \Delta_0 - E)$. 
In this case, 
$\{Q_{-},Q_{+}\} = 
b^2 [{\cal H} - (M+\Delta_0)][{\cal H}- (M+\Delta_0)]$ 
is proportional to a polynomial of ${\cal H}$, again indicating 
a deformation of the conventional SUSY algebra. 
In real nuclei, the relativistic pseudospin symmetry is slightly broken, 
implying $\Delta(r)\neq \Delta_0$ in Eq.~(\ref{pspot}). 
Taking $H_{\kappa}$ as in Eq.~(\ref{hkgen}) and 
$L$ as in Eq.~(\ref{Ldel0}) but with 
$\Delta_0\rightarrow\Delta(r)$, we find that 
$LH_{\kappa_1} - H_{\kappa_2}L = i\,b\,\Delta^{\prime}\sigma_2$. 
Furthermore, $\{Q_{-},Q_{+}\}$ has the same formal form as before, 
but the appearance of $\Delta(r)$ instead of $\Delta_0$ implies 
that the anticommutator is no longer just a polynomial of ${\cal H}$. 

{\it The spin limit} $(\kappa_1+\kappa_2 =-1)$\\
The solutions of Eq.~(\ref{eqvsvs}) with $\omega_{+}=\kappa_1+\kappa_2=-1$ 
require that the difference of the scalar and vector potentials is a 
constant 
\ba
\Sigma(r) = V_{S}(r) - V_{V}(r) = \Sigma_0 ~.
\label{sig0}
\ea
Under such condition the Dirac Hamiltonian is invariant under another SU(2) 
algebra, whose generators are obtained from Eq.~(\ref{Sgen}) by 
interchanging ${\hat s}_{\mu}$ and $\hat {\tilde s}_{\mu}$~\cite{bell75}
\ba
\hat{S}_{\mu} =
\left (
\begin{array}{cc}
{\hat s}_{\mu} & 0\\
0 & \hat {\tilde s}_{\mu}
\end{array}
\right ) ~.
\label{Spgen}
\ea
This relativistic symmetry gives rise to degenerate doublets of 
Dirac states $\Psi_{\kappa_1<0,m}$ and $\Psi_{\kappa_2>0,m}$ with 
$\kappa_1+\kappa_2=-1$, whose upper components have quantum numbers
$(n,\ell,j = \ell + 1/2)$ and 
$(n,\ell,j = \ell - 1/2)$. 
Such spin doublets were argued to be relevant for 
degeneracies observed in heavy-light quark mesons \cite{page01} 
and possibly for the anti-nucleon spectrum in a nucleus \cite{ring03}. 
In the spin limit, 
the corresponding radial components satisfy 
$G_{\kappa_1} = G_{\kappa_2}$ and
$\frac{dF_{\kappa_1}}{dr} - \frac{\kappa_1}{r}F_{\kappa_1} 
= \frac{dF_{\kappa_2}}{dr} - \frac{\kappa_2}{r}F_{\kappa_2}$. 
As shown in Fig.~1(d), the spectrum consists of towers of states with 
$\kappa_1+\kappa_2=-1$, forming pair-wise degenerate spin doublets. 
In this case, none of the towers have a single non-degenerate state 
and hence, 
the spectrum corresponds to that of a broken SUSY \cite{junker96}. 
The tower with $\kappa_1=-1$ ($s_{1/2}$ states) is on its own, since 
states with $\kappa_2=0$ do not exist. The transformation operator, 
found from Eqs.~(\ref{ABmat})-(\ref{eqvsvs}), 
\ba
L = -b \left (
\begin{array}{cc}
(2M + \Sigma_0 + \Delta ) \quad & 
-\frac{d}{dr} + \frac{\kappa_1}{r} \\
\frac{d}{dr} + \frac{\kappa_2}{r} \quad & 0  
\end{array}
\right ) ~,
\label{Lsig0}
\ea 
connects the two doublet states 
as in Eq.~(\ref{LGF}), but with $\kappa_1+\kappa_2=-1$ 
and $C = -b(E + M + \Sigma_0)$. 
The nilpotent charges, $Q_{\pm}$, commute with the supersymmetric 
Hamiltonian, ${\cal H}$, and exhibit a deformation of the conventional 
SUSY algebra, 
$\{Q_{-},Q_{+}\} = 
b^2 [{\cal H} + (M+\Sigma_0)][{\cal H}+ (M+\Sigma_0)]$. 
When $\Sigma(r)\neq \Sigma_0$ in Eq.~(\ref{sig0}), we have  
$LH_{\kappa_1} - H_{\kappa_2}L = -i\,b\,\Sigma^{\prime}\sigma_2$, 
where $H_{\kappa}$ is given in 
Eq.~(\ref{hkgen}) and $L$, as well as $\{Q_{-},Q_{+}\}$, have the 
same form as before 
but with $\Sigma_0\rightarrow\Sigma(r)$. 

In summary, a common intertwining relation was shown to provide the basis 
for a unified treatment of three separate limits at which 
a Dirac Hamiltonian, with scalar and vector potentials, 
exhibits supersymmetric patterns. 
In the Coulomb limit the potentials are $1/r$ but their strengths are 
otherwise arbitrary. In the pseudospin or spin limits there are no 
restrictions on the $r$-dependence of the potentials but there is a 
constraint on their sum or difference. The characteristic degeneracies 
reflect the presence of additional conserved operators, 
the generalized Johnson-Lippmann operator given in Eq.~(\ref{JL}), 
and the previously introduced relativistic pseudospin and spin 
generators \cite{bell75,ginolev98}. 
It is gratifying to note that the indicated supersymmetric patterns 
are manifested empirically, to a good 
approximation, in physical dynamical systems. 
While previous studies have focused on individual doublets in
nuclei and hadrons, it is the grouping of several doublets (and 
intruder levels in nuclei), as suggested in the present work, which
highlights the fingerprints of supersymmetry present
in these dynamical systems.
This work was supported by the Israel Science Foundation.

\clearpage
\begin{figure}
\caption{Schematic qualitative supersymmetric patterns in (a) SUSYQM 
and in the (b) Coulomb, (c) pseudospin, (d) spin, limits 
of the Dirac Hamiltonian. 
In (a) $H_1$ and $H_2$ have identical spectra with an additional level 
for $H_1$ when SUSY is exact. 
Spectroscopic notation $n \ell j$ in (b)-(d) refers 
to quantum numbers of the upper component, and $\kappa$, $N$, $\tl$ 
are defined in the text. 
In (b) the radial nodes $n$ are related to $n_r$ by 
$n_r = n$ $(n_r = n +1)$ for $\kappa<0$ $(\kappa>0)$, and only the 
$E^{(+)}_{n_r,\kappa}$ branch is shown.}
\end{figure}


\newpage

\begin{thebibliography}{99}

\bibitem{wal86}
B.D. Serot and J.D. Walecka, 
Adv. Nucl. Phys. {\bf 16}, 1 (1986). 

\bibitem{gromes91}
W. Lucha, F.F. Sch\"{o}berl and D. Gromes, 
Phys. Rep. {\bf 200}, 127 (1991).

\bibitem{gino97}
J.N. Ginocchio, 
Phys. Rev. Lett. {\bf 78}, 436 (1997). 

\bibitem{page01}
P.R. Page, T. Goldman and J.N. Ginocchio, 
Phys. Rev. Lett. {\bf 86}, 204 (2001). 

\bibitem{Thaller92}
B. Thaller, 
{\it The Dirac Equation}, 
(Springer-Verlag, 1992).

\bibitem{suku85}
C.~V. Sukumar,
J. Phys. A {\bf 18}, L697 (1985).

\bibitem{witten81}
E. Witten, 
Nucl. Phys. B {\bf 188}, 513 (1981).

\bibitem{junker96}
G. Junker, 
{\it Supersymmetric Methods in Quantum and Statistical Physics}, 
(Springer-Verlag, 
Heidelberg 1996).

\bibitem{mul85}
W. Greiner, B. M\"{u}ller and J. Rafelski, 
{\it Quantum Electrodynamics of
Strong Fields} (Springer-Verlag, 
NY, 1985).

\bibitem{deber02}
N. Debergh {\it et al.}, 
J. Phys. A {\bf 35}, 3279 (2002);
L.M. Nieto {\it et al.}, 
Ann. Phys. {\bf 305}, 151 (2003).

\bibitem{john50}
M.~H. Johnson and B.~A. Lippmann, 
Phys. Rev. {\bf 78}, 329 (1950).

\bibitem {bell75}
J.S. Bell and H. Ruegg,
Nucl. Phys. B {\bf 98}, 151 (1975).

\bibitem{ginolev98}
J.N. Ginocchio and  A. Leviatan,
Phys. Lett. B {\bf 425}, 1 (1998).

\bibitem{arima69}
K.T. Hecht and A. Adler,
Nucl. Phys. A {\bf 137}, 129 (1969); 
A. Arima, M. Harvey and K. Shimizu,
Phys. Lett. B {\bf 30}, 517 (1969).

\bibitem{bohr82}
A. Bohr, I. Hamamoto and B.R. Mottelson,
Phys. Scr. {\bf 26}, 267 (1982).

\bibitem{levgino01}
A. Leviatan and J.N. Ginocchio,
Phys. Lett. B {\bf 518}, 214 (2001).

\bibitem{ginomad98}
J.N. Ginocchio and D.G. Madland, 
Phys. Rev. C {\bf 57}, 1167 (1998); 
G.A. Lalazissis {\it et al.}, 
{\it ibid.} {\bf 58}, R45 (1998); 
J. Meng {\it et al.}, 
{\it ibid.} {\bf 59}, 154 (1999). 
 
\bibitem{ginolev01}
J.N. Ginocchio and A. Leviatan,
Phys. Rev. Lett. {\bf 87}, 072502 (2001); 
J.N. Ginocchio,
Phys. Rev. C {\bf 66}, 064312 (2002).

\bibitem{ring03}
S.G. Zhou, J. Meng and P. Ring, 
Phys. Rev. Lett. {\bf 91}, 262501 (2003).

\end{thebibliography}
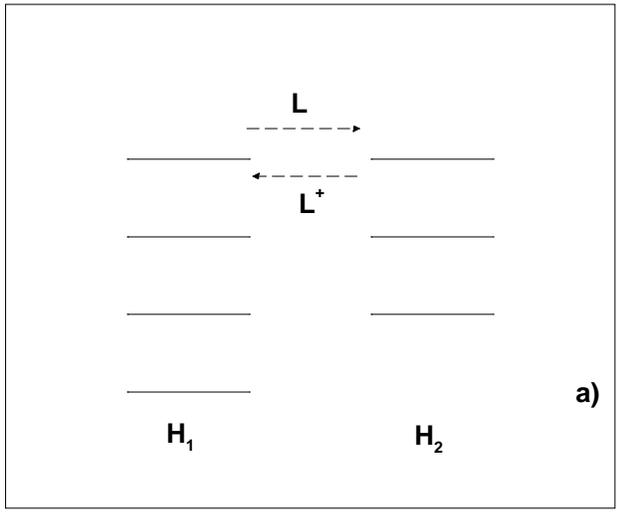
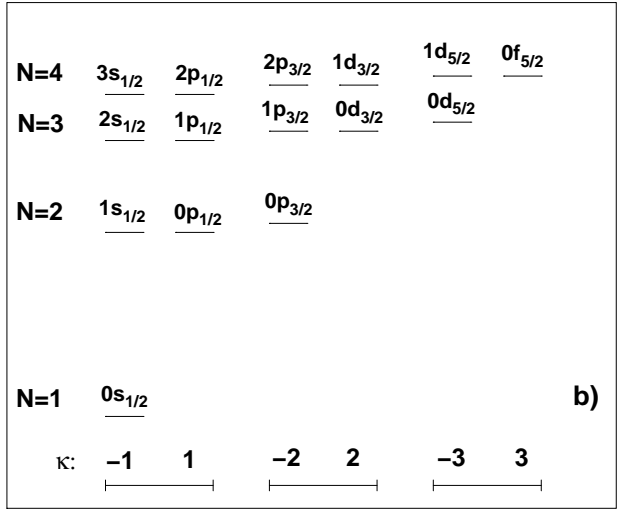
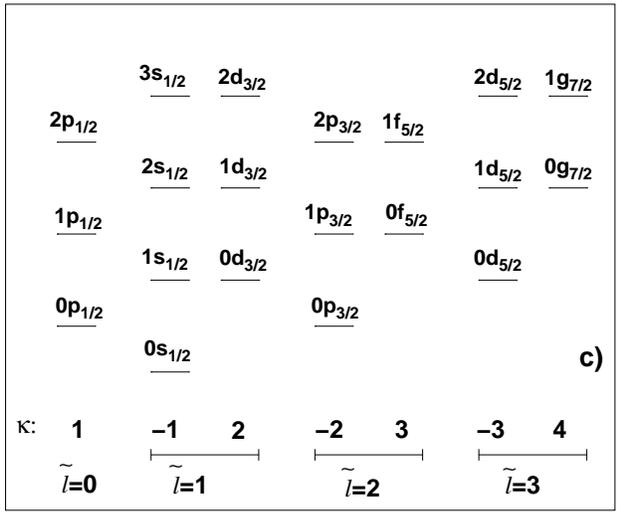
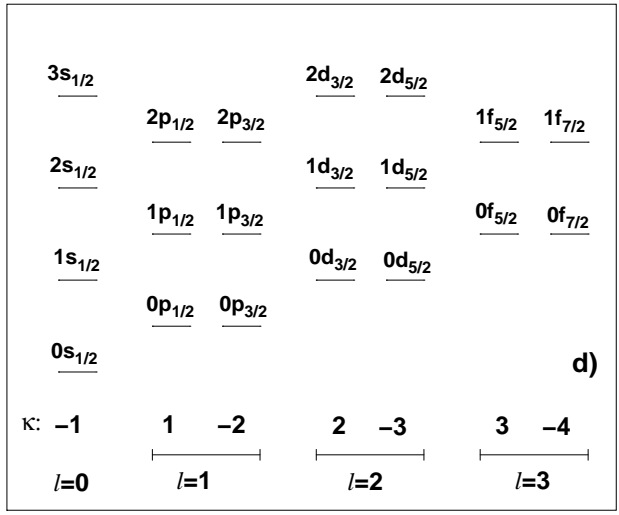
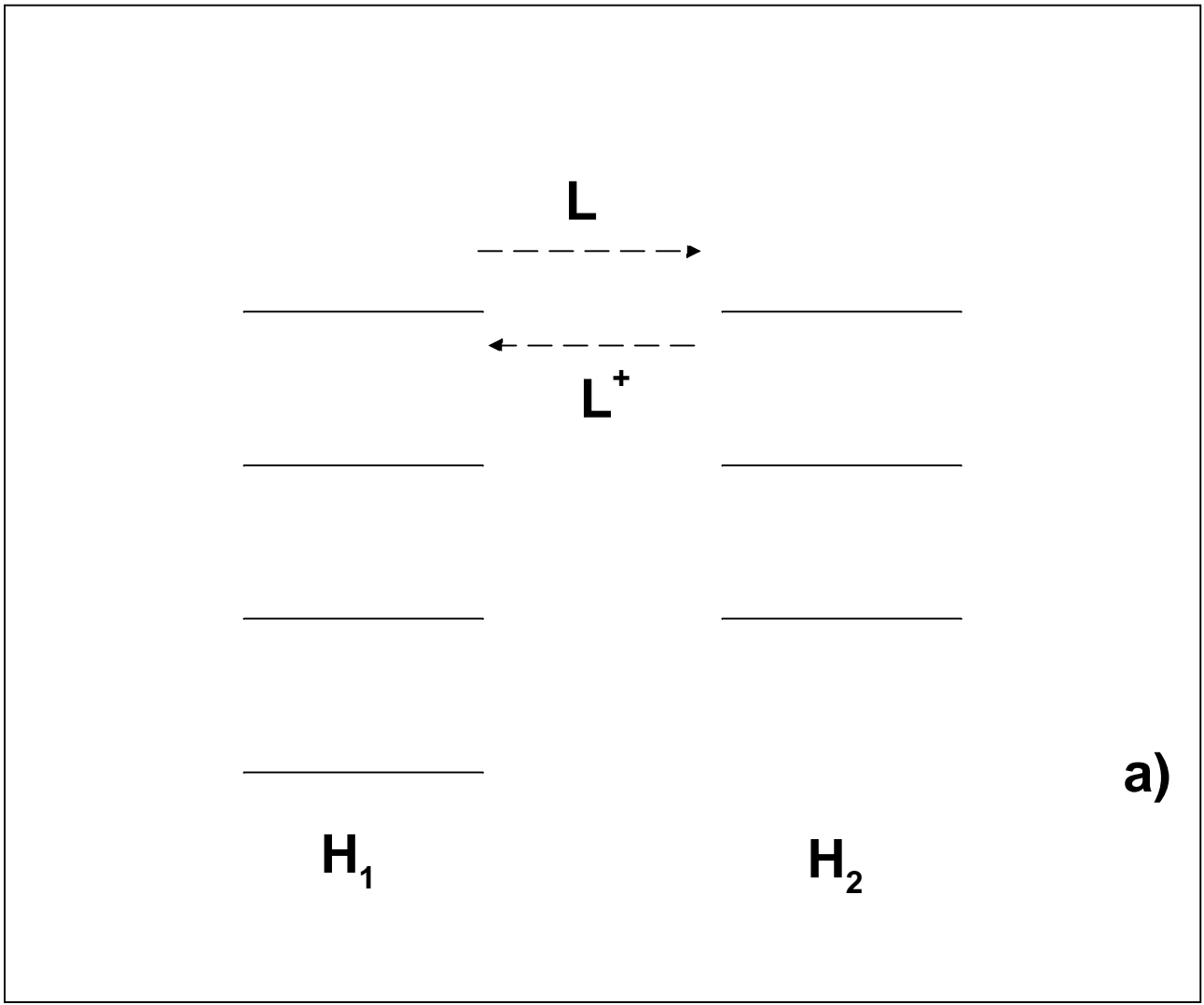
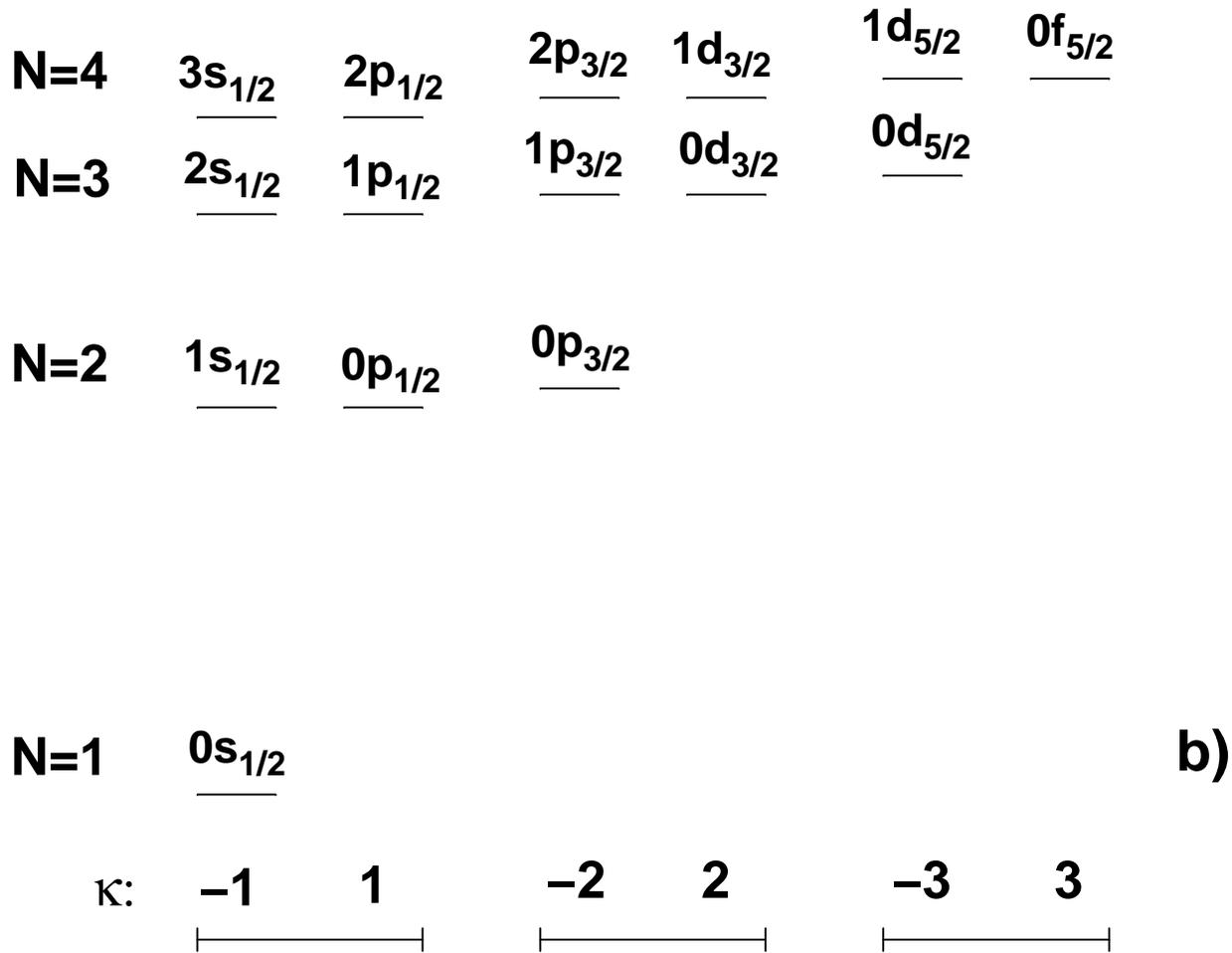
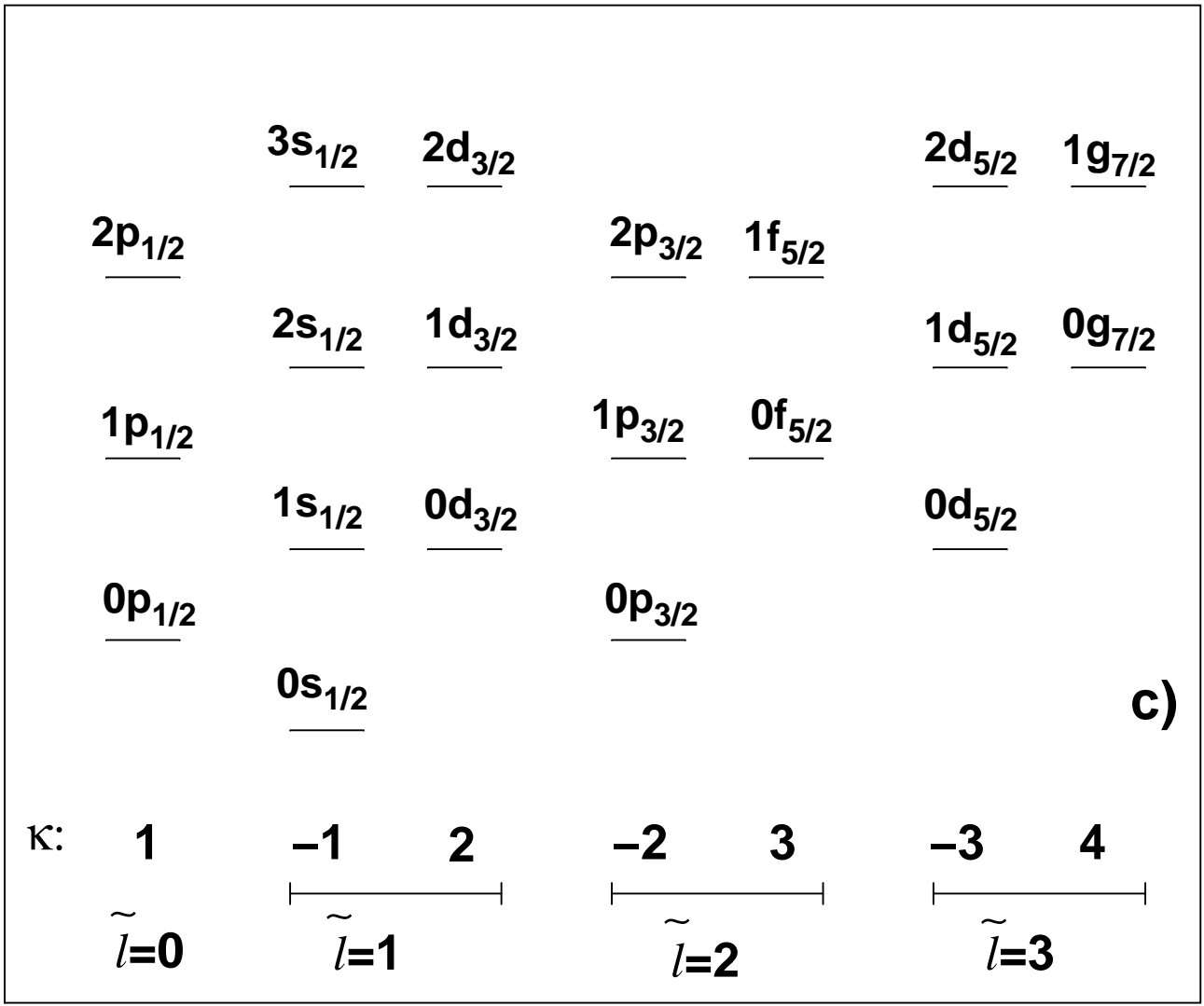
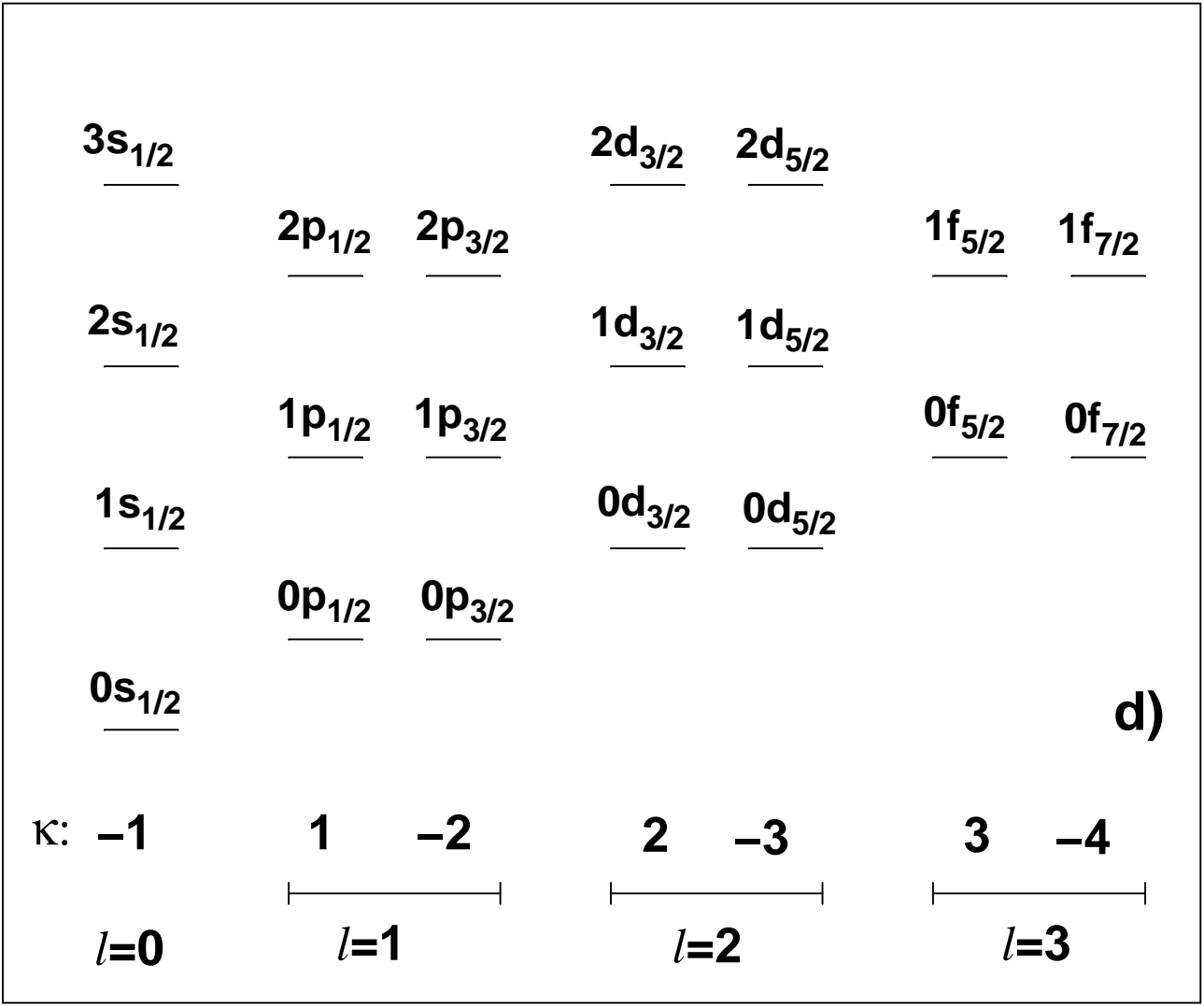
\end{document}